\documentclass[12pt]{article}

\usepackage[T1]{fontenc}
\usepackage[utf8]{inputenc}
\usepackage{lmodern}
\usepackage{amsthm}
\usepackage{amssymb}
\usepackage{amstext}
\usepackage{bm}
\usepackage{graphicx}
\usepackage{epstopdf}
\usepackage[export]{adjustbox}
\usepackage{subcaption}
\usepackage[breaklinks=true]{hyperref}
\usepackage{amsmath}
\usepackage{setspace}
\usepackage{morefloats} 
\usepackage[capitalize]{cleveref}

\crefalias{subequation}{equation}
\crefformat{pluraleq}{Eqs.~(#2#1#3)}

\renewcommand{\baselinestretch}{1.0} 
\makeatletter
\newcounter{parentsubcaption}

\makeatother

\hypersetup{colorlinks=true,
  pdftitle={Simplified PN Equations for Nonclassical Transport with Isotropic Scattering},
  pdfauthor={R. Vasques and R.N. Slaybaugh}
}

\renewcommand{\vec}[1]{\bm{#1}} 
\def\bal#1\nal{\begin{align}#1\end{align}}
\def\bala#1\nala{\begin{align*}#1\end{align*}}
\def\bsub#1\nsub{\begin{subequations}#1\end{subequations}}
\newcommand{\f}{\frac}
\newcommand{\ux}{{\bm x}}
\newcommand{\un}{{\bm n}}
\newcommand{\unab}{{\bf \nabla}}
\newcommand{\bg}{\big>}
\newcommand{\bl}{\big<}
\newcommand{\su}{\big< s\big>}
\newcommand{\sd}{\big< s^2\big>}
\newcommand{\st}{\big< s^3\big>}
\newcommand{\sq}{\big< s^4\big>}
\renewcommand{\sc}{\big< s^5\big>}
\renewcommand{\ss}{\big< s^6\big>}
\newcommand{\sue}{\big< s\big>_\epsilon}
\newcommand{\sde}{\big< s^2\big>_\epsilon}
\newcommand{\ste}{\big< s^3\big>_\epsilon}
\newcommand{\sqe}{\big< s^4\big>_\epsilon}
\newcommand{\sce}{\big< s^5\big>_\epsilon}
\newcommand{\sse}{\big< s^6\big>_\epsilon}
\newcommand{\wsa}{\widehat\Sigma_a}
\newcommand{\wst}{\widehat\Sigma_t}
\newcommand{\wq}{\widehat Q}
\newcommand{\ep}{\varepsilon}
\newcommand{\vi}{{\varphi}}
\newcommand{\uom}{{\bf \Omega}}
\newcommand{\setb}{\mathcal{B}}

\begin{document}

\title{Simplified P$_N$ Equations for Nonclassical Transport with Isotropic Scattering}
\author{{\bf R.\ Vasques}\footnote{Email: \texttt{richard.vasques@fulbrightmail.org}} , {\bf R.N.\ Slaybaugh}\\ \\
\em Department of Nuclear Engineering\\
\em University of California, Berkeley\\
\em Berkeley, CA 94720-1730}
\date{}
\maketitle

\begin{abstract}
\noindent An asymptotic analysis is used to derive a set of diffusion approximations to the nonclassical transport equation with isotropic scattering.
These approximations are shown to reduce to the simplified P$_N$ equations under the assumption of classical transport, and therefore are labeled nonclassical SP$_N$ equations.
In addition, the nonclassical SP$_N$ equations can be manipulated into a classical form with modified parameters, which can be implemented in existing SP$_N$ codes.
Numerical results are presented for an one-dimensional random periodic system, validating the theoretical predictions.
\end{abstract}

\doublespacing

\section{Introduction}

The nonclassical theory of linear particle transport \cite{lar07,larvas11} was developed to address transport problems in which the particle flux is not attenuated exponentially.
This is the case in certain inhomogeneous random media in which the locations of the
scattering centers are spatially correlated.
The nonclassical transport equation consists of a linear Boltzmann equation on
an extended phase space, able to model particle transport for any given free-path
distribution.
Applications of this nonclassical theory include neutron transport in reactor cores (cf. \cite{vaslar14b}), radiative transfer in atmospheric clouds (cf. \cite{davxu14}), and computer graphics (cf. \cite{deon13}).

In this paper we consider the one-speed nonclassical transport equation with isotropic scattering.
This equation is written as
\bal\label{1}
&\f{\partial \Psi}{\partial s}(s) + \uom\cdot\unab\Psi(s) + \Sigma_t(s)\Psi(s) = \f{\delta(s)}{4\pi}\left[\int_{4\pi}\int_0^\infty c\Sigma_t(s')\Psi(\ux,\uom',s')ds'd\Omega' + Q(\ux)\right],
\nal
where $s$ describes the free-path of a particle (distance traveled since the particle's previous interaction), $\Psi(s)=\Psi(\ux,\uom,s)$ is the nonclassical angular flux, $c$ is the scattering ratio (probability of scattering), and $Q$ is an isotropic source.
The total cross section $\Sigma_t$ is a function of the free-path $s$ and satisfies \bal\label{2}
p(s)= \Sigma_t(s)e^{-\int_0^s\Sigma_t(s')ds'},
\nal
where $p(s)$ is the free-path distribution function.

The particle flux in its standard definition can be recovered from the solution of \cref{1} by integrating over the free-path $s$, such that
\bsub\label[pluraleq]{3}
\bal\label{3a}
\Psi_c(\ux,\uom) = \int_0^\infty\Psi(\ux,\uom,s)ds = \text{classical angular flux,}\nal
and 
\bal\label{3b}
\Phi(\ux) = \int_{4\pi}\int_0^\infty\Psi(\ux,\uom,s)dsd\Omega= \text{scalar flux}.
\nal
\nsub

For $m=0, 1, 2,...,$ we define the $m$-th raw moment of $p(s)$ as
\bsub\label[pluraleq]{4}
\bal\label{4a}
\bl s^m\bg = \int_0^\infty s^mp(s)ds.
\nal
The following identity holds for $m=1,2,...$ : 
\bal\label{4b}
\bl s^m\bg = m\int_0^\infty s^{m-1}e^{-\int_0^s\Sigma_t(s')ds'}ds.
\nal
\nsub
Assuming $\sd < \infty$, an asymptotic approximation of \cref{1} for the scalar flux given in \cref{3b} has been formally derived \cite{lar07,larvas11}:
\bal\label{5}
-\f{1}{6}\f{\sd}{\su}\nabla^2\Phi(\ux) + \f{1-c}{\su}\Phi(\ux) = Q(\ux).
\nal
Convergence of \cref{1} to the \textit{nonclassical diffusion equation} (\cref{5}) has been rigorously discussed in \cite{fragou10}.

If the free-path distribution is given by the exponential $p(s) = \Sigma_te^{-\Sigma_ts}$, the raw moments defined in \cref{4} yield
\bal\label{6}
\bl s^m\bg = \int_0^\infty s^m\Sigma_te^{-\Sigma_t s}ds = \f{m!}{\Sigma_t^m}.
\nal
In this situation, \cref{1} reduces to the classical transport equation
\bsub\label[pluraleq]{7}
\bal\label{7a}
\uom\cdot\unab\Psi_c(\ux,\uom)+\Sigma_t\Psi_c(\ux,\uom) = \f{1}{4\pi}\left[c\Sigma_t\Phi(\ux) + Q(\ux)\right],
\nal
and \cref{5} reduces to
\bal\label{7b}
-\f{1}{3\Sigma_t}\nabla^2\Phi(\ux) + \Sigma_a\Phi(\ux) = Q(\ux),
\nal
\nsub
where $\Sigma_a = (1-c)\Sigma_t$ is the absorption cross section.

The classical diffusion equation (\ref{7b}) has been generalized to the hierarchy of the simplified P$_N$ (SP$_N$) equations, first derived by Gelbard \cite{gel60,gel61,gel62}.
These equations were shown to be a high-order asymptotic approximation of the transport equation \cite{larmor93}.
We refer the reader to \cite{mcl11} for a complete review on SP$_N$ theory.

In this paper we use an asymptotic analysis to derive more accurate diffusion approximations to \cref{1}.
We show that, if $p(s)$ is given by an exponential (classical transport), these approximations reduce to the classical SP$_N$ equations; therefore, they are labeled \textit{nonclassical} SP$_N$ equations.

The remainder of this paper is organized as follows.
The asymptotic analysis is carried out in Section II, in which we also provide explicit formulations for the nonclassical SP$_1$ (diffusion), SP$_2$, and SP$_3$ equations. 
Nonclassical SP$_N$ equations for $N > 3$ can be derived by continuing the same procedure.
In Section III we show that if \cref{6} holds (classical transport), the nonclassical SP$_N$ equations reduce to the classical SP$_N$ equations.
In Section IV we show that the nonclassical SP$_N$ equations can be manipulated into a classical form with modified parameters, allowing the use of classical Marshak boundary conditions.  
Section V describes numerical results that validate the theoretical predictions.
We conclude with a brief discussion in Section VI.

\section{Asymptotic Analysis}\label{sec2}
Let us write \cref{1} in the mathematically equivalent form
\bsub\label[pluraleq]{8}
\bal
&\f{\partial \Psi}{\partial s}(s) + \uom \cdot \unab \Psi(s) + \Sigma_t(s)\Psi(s) = 0, \qquad s>0,\\
&\Psi(0) = \f{1}{4\pi}\left[\int_{4\pi}\int_0^\infty c\Sigma_t(s')\Psi(\ux,\uom',s')ds'd\Omega' + Q(\ux)\right],
\nal
\nsub
where $\Psi(0) = \displaystyle{\lim_{s\to 0^+}\Psi(s)} = \Psi(0^+)$.
Defining $0<\ep\ll 1$, we perform the following scaling:
\bsub\label[pluraleq]{9}
\bal
\Sigma_t(s) &= \f{\Sigma_t(s/\ep)}{\ep},\label{9a}\\
c &= 1-\ep^2\kappa,\label{9b}\\
Q(\ux) &= \ep q(\ux)\label{9c},
\nal
where $\kappa$ and $q$ are $O(1)$.
\Cref{9a,9b,9c} are equivalent to the scaling used in \cite{larmor93} to obtain the classical SP$_N$ approximations for \cref{7a}.
Moreover, using \cref{2,4a,9a}, we can define $\bl s^m\bg_\ep$ such that
\bal\label{9d}
\bl s^m\bg &=
\ep^m\int_0^\infty \left(\f{s}{\ep}\right)^m \f{1}{\ep}\Sigma_t(s/\ep)e^{-\int_0^s\f{1}{\ep}\Sigma_t(s'/\ep)ds'}ds  \\
&=
\ep^m\int_0^\infty s^m\Sigma_t(s)e^{-\int_0^s\Sigma_t(s')ds'}ds\nonumber\\
&= \ep^m\bl s^m\bg_\ep\nonumber,
\nal
\nsub
where $\bl s^m\bg_\ep$ is $O(1)$. This scaling implies that:
\begin{itemize}
\item The system is optically thick.\vspace{-5pt}
\item The transport process is dominated by scattering, described by the terms of $O(\ep^{-1})$.\vspace{-5pt}
\item Absorption and source are small and comparable [$O(\ep)$].\vspace{-5pt}
\item Both the infinite medium solution $\Phi=Q/\Sigma_a$ and the diffusion length $(3\Sigma_t\Sigma_a)^{-1/2}$ are $O(1)$.\vspace{-5pt}
\item The equations for nonclassical (\cref{5}) and classical (\cref{7b}) diffusion are $\ep$-invariant.
\end{itemize}

With this scaling, \cref{8} become
\bala
&\f{\partial \Psi}{\partial s}(s) + \uom \cdot \unab \Psi(s) + \f{1}{\ep}\Sigma_t(s/\ep)\Psi(s) = 0, \qquad s>0,\\
&\Psi(0) = \f{1}{4\pi}\left[\int_{4\pi}\int_0^\infty \f{(1-\ep^2\kappa)}{\ep}\Sigma_t(s'/\ep)\Psi(\ux,\uom',s')ds'd\Omega' + \ep q(\ux)\right].
\nala
Next, we define
\bala
\Psi(\ux,\uom,\ep s) \equiv \Psi_\ep(\ux,\uom,s),
\nala
which satisfies
\bala
&\f{\partial \Psi_\ep}{\partial s}(s) + \ep\uom \cdot \unab \Psi_\ep(s) + \Sigma_t(s)\Psi_\ep(s) = 0, \qquad s>0,\\
&\Psi_\ep(0) = \f{1}{4\pi}\left[\int_{4\pi}\int_0^\infty (1-\ep^2\kappa)\Sigma_t(s')\Psi_\ep(\ux,\uom',s')ds'd\Omega' + \ep q(\ux)\right].
\nala
Then, defining
\bala
\Psi_\ep(\ux,\uom,s) \equiv \psi(\ux,\uom,s)\f{e^{-\int_0^s\Sigma_t(s')ds'}}{\ep\sue},
\nala
where $\psi(\ux,\uom,s)$ satisfies
\bsub\label[pluraleq]{10}
\bal
&\f{\partial \psi}{\partial s}(s) + \ep\uom \cdot \unab \psi(s)= 0, \qquad s>0,\label{10a}\\
&\psi(0) = \label{10b} \f{1}{4\pi}\left[\int_{4\pi}\int_0^\infty (1-\ep^2\kappa)p(s')\psi(\ux,\uom',s')ds'd\Omega' + \ep^2 \sue q(\ux)\right].
\nal
\nsub
We remark that the scalar flux defined in \cref{3b} can be written as
\bal\label{11}
\Phi(\ux) &= \int_{4\pi}\int_0^\infty \ep\Psi_\ep(\ux,\uom,s) dsd\Omega \\
& = \int_{4\pi}\int_0^\infty \psi(\ux,\uom,s)\f{e^{-\int_0^s\Sigma_t(s')ds'}}{\sue}ds d\Omega.\nonumber
\nal

Integrating \cref{10a} over $0<s'<s$ and using \cref{10b}, we obtain
\bal\label{12}
&\left(I+\ep\uom\cdot\unab\int_0^s(\cdot)ds\right)\psi = \f{1}{4\pi}\left[\int_0^\infty (1-\ep^2\kappa)p(s')\vi(\ux,s')ds' + \ep^2\sue q\right],
\nal
where
\bala
\vi(\ux,s) = \int_{4\pi}\psi(\ux,\uom,s)d\Omega.
\nala
Inverting the operator on the left-hand side of \cref{12} and expanding it in a power series, we obtain
\bal\label{13}
\psi &= \left(\sum_{n=0}^{\infty}(-\ep)^n\left(\uom\cdot\unab\int_0^s(\cdot)ds\right)^n\right) \f{1}{4\pi}\left[\int_0^\infty (1-\ep^2\kappa)p(s')\vi(\ux,s')ds' + \ep^2\sue  q\right].
\nal

Let us define
\bsub\label[pluraleq]{14}
\bal
\unab_0 &= \f{1}{3}\unab^2,\label{14a}\\
\setb &=\unab_0\left(\int_0^s(\cdot)ds\right)^2.\label{14b}
\nal
\nsub
Then, using the identity \cite{fra07}
\bala
\f{1}{4\pi}\int_{4\pi}\left(\uom\cdot\unab\int_0^s(\cdot)ds\right)^nd\Omega = 
\f{1+(-1)^n}{2}\f{(3\setb)^{n/2}}{n+1},
\nala
for $n=0,1,2,...$ , we integrate \cref{13} over the unit sphere and obtain
\bala
\vi &= \left(\sum_{n=0}^{\infty}\f{1}{2n+1}(3\ep^2\setb)^n\right)\left[\int_0^\infty (1-\ep^2\kappa)p(s')\vi(\ux,s')ds' + \ep^2\sue q\right].
\nala
Inverting the operator on the right-hand side of this equation and once again expanding it in a power series, we get
\bal
&\left(I- \ep^2 \setb - \f{4\ep^4}{5}\setb^2 - \f{44\ep^6}{35}\setb^3 + O(\ep^8)\right)\vi =\label{15}\int_0^\infty (1-\ep^2\kappa)p(s')\vi(\ux,s')ds' + \ep^2\sue q.
\nal
The solution of this equation is
\bal\label{16}
&\vi(\ux,s) = \left(I + \ep^2\f{s^2}{2!}\unab_0 + \f{9\ep^4}{5}\f{s^4}{4!}\unab_0^2 + \f{27\ep^6}{7}\f{s^6}{6!}\unab_0^3 + O(\ep^8)\right)\phi(\ux),
\nal
where
\bala
\phi(\ux) = {\sum_{n=0}^\infty  \ep^{2n}\phi_{2n}(\ux)},
\nala
with $\phi_{2n}(\ux)$ undetermined at this point.

We multiply \cref{16} by $e^{-\int_0^s\Sigma_t(s')ds'}/\sue$ and operate on it by $\int_0^\infty (\cdot) ds$.
Using \cref{4b,9d,11}, we obtain an expression for the scalar flux:
\bala
\Phi(\ux) = &\left(I + \ep^2\f{\ste}{3!\sue}\unab_0 + \f{9\ep^4}{5}\f{\sce}{5!\sue}\unab_0^2 + \f{27\ep^6}{7}\f{\bl s^7\bg_\ep}{7!\sue}\unab_0^3 + O(\ep^8)\right)\phi(\ux). 
\nala
Hence, we can write
\bal\label{17}
\int_0^\infty p(s)\vi(\ux,s)ds = \left(\sum_{n=0}^\infty\ep^{2n}U_n\unab_0^n \right)\Phi(\ux),
\nal
with
\bala
U_0 &= 1,\\
U_1&= \f{\sde}{2!}-\f{\ste}{3!\sue},\\
U_2 &= \f{9}{5}\left[\f{\sqe}{4!}-\f{\sce}{5!\sue}\right]-\f{\ste}{3!\sue}U_1,\\
U_3 &=\f{27}{7}\left[\f{\sse}{6!}-\f{\bl s^7\bg_\ep}{7!\sue}\right]-\f{9}{5}\f{\sce}{5!\sue}U_1-\f{\ste}{3!\sue}U_2,\\
& \,\,\, \vdots  
\nala
\Cref{15} can be rewritten as
\bal\label{18}
&\left(\sum_{n=0}^\infty \ep^{2n}V_n\unab_0^n\right)\Phi(\ux) =  (1-\ep^2\kappa)\left(\sum_{n=0}^\infty\ep^{2n}U_n\unab_0^n \right)\Phi(\ux) + \ep^2\sue q(\ux),
\nal
where
\bala
V_0 &= 1,\\
V_1&= -\f{\ste}{3!\sue}V_0,\\
V_2 &= -\f{9}{5}\f{\sce}{5!\sue}V_0 -\f{\ste}{3!\sue}V_1,\\
V_3 &= -\f{27}{7}\f{\bl s^7 \bg_\ep}{7!\sue}V_0 - \f{9}{5}\f{\sce}{5!\sue}V_1-\f{\ste}{3!\sue}V_2,\\
& \,\,\, \vdots 
\nala

Finally, rearranging the terms in \cref{18} we get
\bal\label{19}
\left(\sum_{n=0}^\infty \ep^{2n}\left[W_{n+1}\unab_0^{n+1}+\kappa U_n\unab_0^n\right]\right)\Phi(\ux) = \sue q(\ux),
\nal
where $W_{n}=V_n-U_n$.
If we discard the terms of $O(\ep^{2n})$ in this equation, we obtain a partial differential equation for $\Phi(\ux)$ of order $2n$.
We will use this approach to explicitly derive the nonclassical SP$_1$, SP$_2$, and SP$_3$ equations.
Higher-order equations can be derived from \cref{19} by continuing to follow the same procedure.

We note that the asymptotic analysis presented in this section requires the first $2M$ raw moments of $p(s)$ to exist in order to obtain the nonclassical SP$_N$ equations for $N=M$.
Specifically, if $p(s)$ decays algebraically as $s\rightarrow\infty$ such that
\bala
p(s)\geq\f{\text{constant}}{s^{2M+1}}\quad  \text {for $s\gg 1$},
\nala
then
\bala
\bl s^{2M}\bg = \int_0^\infty s^{2M}p(s)ds = \infty,
\nala
and the asymptotic theory developed above is invalid.
In particular, the case of $\sd = \infty$ (known as ``anomalous'' or ``generalized'' diffusion) is relevant to several radiative transfer problems in atmospheric sciences \cite{davxu14}.

\subsection{Nonclassical Diffusion Equation (Nonclassical SP$_1$)}
We discard the terms of $O(\ep^2)$ in \cref{19} and rewrite the equation as 
\bala
W_1\unab_0\Phi(\ux) + \kappa\Phi(\ux) = \sue q(\ux).
\nala
Using \cref{14a}, we get
\bala
- \f{1}{6}\f{\sde}{\sue}\unab^2\Phi(\ux) + \f{\kappa}{\sue}\Phi(\ux) = q(\ux).
\nala
Multiplying this equation by $\ep$ and using \cref{9} to revert to the original unscaled parameters, we obtain
\bal\label{20}
- \f{1}{6}\f{\sd}{\su}\unab^2\Phi(\ux) + \f{1-c}{\su}\Phi(\ux) = Q(\ux),
\nal
which is the nonclassical diffusion equation (\ref{5}) as derived in \cite{lar07,larvas11}. 

\subsection{Nonclassical simplified P$_2$ equation}
Discarding the terms of $O(\ep^4)$ in \cref{19}, we have
\bala
\left(W_1\unab_0 + \ep^2\left[W_2\unab_0^2 + \kappa U_1\unab_0\right]\right)	\Phi(\ux) + \kappa\Phi(\ux) = \sue q(\ux).
\nala
We rearrange the terms of this equation to get
\bala
-\left(I + \ep^2\f{W_2\unab_0 +\kappa U_1}{W_1}\right)W_1\unab_0\Phi(\ux) = \kappa\Phi(\ux) - \sue q(\ux).
\nala
Operating on this equation by $\left(I - \ep^2\left[W_2\unab_0+\kappa U_1\right]/W_1\right)$ and discarding terms of $O(\ep^4)$, it becomes
\bala
W_1\unab_0& \left[\Phi(\ux)-\ep^2\f{W_2}{W_1^2}\left[\kappa\Phi(\ux) - \sue q(\ux)\right]\right] + \\
&\qquad \kappa \left[1-\ep^2\kappa\f{U_1}{W_1}\right]\Phi(\ux) =
\left[1-\ep^2\kappa\f{U_1}{W_1}\right]\sue q(\ux).
\nonumber
\nala

Finally, we multiply this equation by $\ep$ and use \cref{9} to revert to the original unscaled parameters. 
Using \cref{14a}, we obtain the \textit{nonclassical SP$_2$ equation}
\bal\label{21}
-\f{1}{6}\f{\sd}{\su}\unab^2 &\bigg[\Phi(\ux)+\lambda_1\left[(1-c)\Phi(\ux) - \su Q(\ux)\right]\bigg]+\\
 &\f{1-c}{\su} \big[1-\beta_1(1-c)\big]\Phi(\ux) =
\big[1-\beta_1(1-c)\big]Q(\ux),\nonumber
\nal
where the constants
\bsub\label[pluraleq]{22}
\bal
\lambda_1 &= \f{3}{10}\f{\sq}{\sd^2} - \f{1}{3}\f{\st}{\su\sd}\label{22a}
\nal
and
\bal
\beta_1 &= \f{1}{3}\f{\st}{\su\sd} - 1\label{22b}
\nal
\nsub
are both $O(1)$.

\subsection{Nonclassical simplified P$_3$ equations}
Discarding the terms of $O(\ep^6)$ in \cref{19}, we have
\bal\label{23}
&\left(W_1\unab_0 + \ep^2\left[W_2\unab_0^2 + \kappa U_1\unab_0\right]+ \ep^4\left[W_3\unab_0^3 + \kappa U_2\unab_0^2\right]\right)\Phi(\ux) + \kappa\Phi(\ux) = \sue q(\ux).
\nal
We define
\bal\label{24}
\nu(\ux) &= \left(\f{\ep^2}{2}\f{W_2}{W_1}\unab_0 +\f{\ep^4}{2}\f{W_3\unab_0^2+\kappa U_2\unab_0}{W_1}\right)\Phi(\ux)\\
&= \left(I + \ep^2\f{W_3\unab_0+\kappa U_2}{W_2}\right)\f{\ep^2}{2}\f{W_2}{W_1}\unab_0\Phi(\ux),\nonumber
\nal
and rewrite \cref{23} as
\bal\label{25}
W_1\unab_0\left[\Phi(\ux) + 2\nu(\ux) + \ep^2\kappa \f{U_1}{W_1}\Phi(\ux)\right] + \kappa\Phi(\ux) = \sue q(\ux).
\nal
Operating on \cref{24} by $\left(I - \ep^2[W_3\unab_0+\kappa U_2]/W_2\right)$ and discarding terms of $O(\ep^6)$, we get
\bala
-\ep^2\unab_0\left[\f{W_3}{W_2}\nu(\ux)+\f{1}{2}\f{W_2}{W_1}\Phi(\ux)\right]+\left[1-\ep^2\kappa\f{U_2}{W_2}\right]\nu(\ux) = 0.
\nala
This equation can be rewritten as
\bal\label{26}
-\ep^2 W_1\unab_0 & \left[\f{W_3}{W_1W_2}\nu(\ux)+\f{1}{2}\f{W_2}{W_1^2}\Phi(\ux)\right]+ \left[1-\ep^2\kappa\f{U_2}{W_2}\right]\nu(\ux) = 0.
\nal

Multiplying \cref{25} by $\ep$ and using \cref{9,14a}, we obtain
\bsub\label[pluraleq]{27}
\bal
&-\f{1}{6}\f{\sd}{\su}\unab^2\bigg[\big[1+\beta_1(1-c)\big]\Phi(\ux) + 2\nu(\ux)\bigg] + \label{27a}\f{1-c}{\su}\Phi(\ux) = Q(\ux),
\nal
where $\beta_1$ is given by \cref{22b}.
Similarly, dividing \cref{26} by $\su$ and using \cref{9,14a}, we obtain
\bal
&-\f{1}{6}\f{\sd}{\su}\unab^2\left[\f{\lambda_1}{2}\Phi(\ux)+\lambda_2\nu(\ux)\right]+\f{1-\beta_2(1-c)}{\su}\nu(\ux) = 0,\label{27b}
\nal
\nsub
where $\lambda_1$ is given by \cref{22a}, and the constants 
\bsub\label[pluraleq]{28}
\bal
\lambda_2 =& \f{1}{10\sd\st-9\su\sq }\left[\f{9}{5}\sc - \f{27}{21}\f{\su\ss}{\sd} + 3\f{\st\sq}{\sd} - \f{10}{3} \f{\st^2}{\su}\right]
\nal
and
\bal
\beta_2 =& \f{1}{10\sde\ste-9\sue\sqe}\left[\f{10}{3}\f{\ste^2}{\sue} - \f{9}{5}\sce\right]-1
\nal
\nsub
are both $O(1)$. \Cref{27} are the \textit{nonclassical SP$_3$ equations}.

\section{Reduction to Classical Theory}
We will show that, in the case of classical transport, the nonclassical SP$_N$ equations derived in the previous section reduce to the SP$_N$ approximations to the classical transport equation (\ref{7a}).
In other words, we now assume that
\bala
\Sigma_t(s) = \Sigma_t \equiv \text{constant (independent of $s$)}.
\nala
Under this assumption, the free-path distribution $p(s)$ is an exponential and \cref{6} holds, such that $\bl s^m\bg = m!\Sigma_t^{-m}$.

Introducing this result into the nonclassical diffusion approximation given by \cref{20}, one can easily see that it reduces to the classical diffusion equation (\ref{7b}).
Moreover, \cref{22} and \cref{28} yield
\bala
\lambda_1 &= \f{4}{5}, \\
\lambda_2 &= \f{11}{7},\\
\beta_1 &= \beta_2 = 0.
\nala
In this case, the nonclassical SP$_2$ equation (\ref{21}) reduces to
\bala
- \f{1}{3\Sigma_t}\unab^2&\left[\Phi(\ux) + \f{4}{5}\f{\Sigma_a\Phi(\ux)- Q(\ux)}{\Sigma_t}\right] + \Sigma_a\Phi(\ux) = Q(\ux),
\nala
which is the classical SP$_2$ approximation to \cref{7a} \cite{larmor93,mcl11}.
The nonclassical SP$_3$ equations (\cref{27}) reduce to 
\bala
-\f{1}{3\Sigma_t}\unab^2\bigg[\Phi(\ux) + 2\nu(\ux)\bigg] + \Sigma_a\Phi(\ux) = Q(\ux),\\
-\f{1}{3\Sigma_t}\unab^2\left[\f{2}{5}\Phi(\ux)+\f{11}{7}\nu(\ux)\right]+\Sigma_t\nu(\ux) = 0,
\nala
which are the classical SP$_3$ approximations to \cref{7a} \cite{larmor93,mcl11}.

Furthermore, if \cref{6} holds, then
\bala
U_1=U_2=U_3 = ... =0,
\nala
and the integral in \cref{17} yields $\Phi(\ux)$.
Defining $\sigma_t = \Sigma_t/\ep$, the terms $W_n$ in the operator on the left side of \cref{19} are
\bala
W_0 &= 1,\\
W_1 &= -\f{1}{\sigma_t^2},\\
W_2 &= -\f{4}{5\sigma_t^4}, \\
W_3 &= -\f{44}{35\sigma_t^6},\\
&\,\,\,\vdots
\nala
Thus, \cref{19} becomes
\bala
-\left(\f{1}{\sigma_t}\unab_0 + \f{4\ep^2}{5\sigma_t^3}\unab_0^2 + \f{44\ep^4}{35\sigma_t^5}\unab_0^3 + O(\ep^6)\right)\Phi(\ux) + \kappa\Phi(\ux) = q(\ux).
\nala
This is the general expression for the asymptotic approximation to \cref{7a} that can be used to obtain the classical SP$_N$ equations \cite{larmor93,mcl11}. 

\section{Boundary Conditions}
The asymptotic analysis presented in this paper does not yield boundary conditions.
To overcome this obstacle, we will show that the nonclassical SP$_N$ equations can be manipulated into a classical form with modified parameters.
This allows the use of classical (Marshak) vacuum boundary conditions \cite{mcl11}.
Moreover, this approach shows that the nonclassical SP$_N$ equations can be implemented in existing SP$_N$ codes with minimal effort.  

\subsection{SP$_1$ Boundary Conditions}
Let us define
\bala
\wst &= 2\f{\su}{\sd}, \\
\wsa &= \f{1-c}{\su}.
\nala
Then, the nonclassical SP$_1$ equation (\cref{20}) can be written in a classical form:
\bsub\label[pluraleq]{29}
\bal
- \f{1}{3\wst}\unab^2\Phi(\ux) + \wsa\Phi(\ux) = Q(\ux).
\nal
The vacuum boundary conditions for this equation are given by 
\bal
\f{1}{2}\Phi(\ux) -\f{1}{3\wst}\vec\un\cdot\unab\Phi(\ux) = 0.
\nal
\nsub
We note that, if \cref{6} holds, $\wst = \Sigma_t$, $\wsa=\Sigma_a$, and \cref{29} represent the classical diffusion equation with Marshak vacuum boundary conditions.

\subsection{SP$_2$ Boundary Conditions}
We define
\bala
\wst &= 2\f{\su}{\sd}, \\
\wsa &=\f{(1-c)}{\su} \f{1-\beta_1(1-c)}{1+\lambda_1(1-c)},\\
\wq(\ux) &= \f{1-\beta_1(1-c)}{1+\lambda_1(1-c)} Q(\ux),\\
\widehat\Phi(\ux) &= \Phi(\ux) + \lambda_1\left[(1-c)\Phi(\ux) - \su Q(\ux)\right].\nala
Then, the nonclassical SP$_2$ equation (\cref{21}) can be manipulated into a classical SP$_2$ equation for the modified flux $\widehat\Phi(\ux)$:
\bsub\label[pluraleq]{30}
\bal
- \f{1}{3\wst}\unab^2\widehat\Phi(\ux) + \wsa\widehat\Phi(\ux) &= \wq(\ux).
\nal
The vacuum boundary conditions for this equation are given by
\bal
\f{1}{2}\widehat\Phi(\ux) -\f{1}{3\wst}\vec\un\cdot\unab\widehat\Phi(\ux) &= 0.
\nal
\nsub
Finally, the scalar flux $\Phi(\ux)$ can be recovered from the solution of \cref{30} using the identity
\bal\label{31}
\Phi(\ux) &= \frac{\widehat\Phi(\ux) + \lambda_1\su Q(\ux)}{1+\lambda_1(1-c)}.
\nal
If \cref{6} holds, \cref{30,31} represent the diffusion form of the classical SP$_2$ equations with Marshak boundary conditions, as described in \cite{tomlar96}.

\subsection{SP$_3$ Boundary Conditions}
We define
\bala
\wst &= 2\f{\su}{\sd},\\
\wsa &=\f{(1-c)}{\su} \f{1}{1+\beta_1(1-c)},\\
\widehat\Sigma_2 &= \f{4\left[1+\beta_1(1-c)\right]\left[1-\beta_2(1-c)\right]}{5\lambda_1\su},\\
\widehat\Sigma_3 &= \f{27}{28}\f{\lambda_1\wst}{\lambda_2\left[1+\beta_1(1-c)\right]-\lambda_1},\\
\wq(\ux) &= \f{Q(\ux)}{1+\beta_1(1-c)},\\
\widehat\Phi_2(\ux) &= \f{\nu(\ux)}{1+\beta_1(1-c)}.
\nala
Then, the nonclassical SP$_3$ equations (\cref{27}) can be manipulated into classical SP$_3$ equations for $\Phi(\ux)$ and $\widehat\Phi_2(\ux)$:
\bsub\label[pluraleq]{32}
\bal
&-\f{1}{3\wst}\unab^2\big[\Phi(\ux) + 2\widehat\Phi_2(\ux)\big] + \wsa\Phi(\ux) = \wq(\ux),\\
&-\f{1}{3\wst}\unab^2\left[\f{2}{5}\Phi(\ux)+\left(\f{4}{5}+\f{27\wst}{35\widehat\Sigma_3}\right)\widehat\Phi_2(\ux)\right]+\widehat\Sigma_2\widehat\Phi_2(\ux) = 0.
\nal
The vacuum boundary conditions for these equations are given by
\bal
&\f{1}{2}\Phi(\ux)-\f{1}{3\wst}\vec\un\cdot\unab\Phi(\ux) -\f{2}{3\wst}\vec\un\cdot\unab\widehat\Phi_2(\ux) +\f{5}{8}\widehat\Phi_2(\ux) = 0,\\
&-\f{1}{8}\Phi(\ux) +\f{5}{8}\widehat\Phi_2(\ux) -\f{3}{7\widehat\Sigma_3} \vec\un\cdot\unab\widehat\Phi_2(\ux) = 0.
\nal
\nsub
As in the previous cases, if \cref{6} holds, $\wst = \widehat\Sigma_2 = \widehat\Sigma_3 =\Sigma_t$, $\wsa=\Sigma_a$, $\wq = Q$, $\widehat\Phi_2 = \nu(\ux)$, and \cref{32} represent the classical SP$_3$ equations with Marshak vacuum boundary conditions.

\section{Numerical Results}
Given the challenges of obtaining benchmark results for multi-dimensional \textit{nonclassical} systems, in which the free-path distribution $p(s)$ is \textit{not} given by an exponential, we will leave that task for the future. 
Numerical results in this paper consider \textit{slab geometry} transport taking place in a one-dimensional (1-D) random periodic system similar to the one introduced in \cite{zuc94}.

The system is formed by a random segment of periodically arranged layers of two materials.
Material 1 is assumed to be highly-scattering, while material 2 is a void.
We note that material 2 being a void does not violate any of our physical assumptions and it corresponds to well-known physical applications \cite{larvas11,vaslar14b,davxu14}.

The ensemble-averaged free-path distribution $p(\mu,s)$ for this 1-D system has been analytically obtained in \cite{vas16}.
Due to the layered nature of the slab, the free-path distribution is angular-dependent; the distance that a particle travels in each layer will depend on its angle of flight, characterized by its cosine $\mu$.
However, the asymptotic analysis in this paper \textit{does not} account for angular-dependent free-path distributions.
Therefore, in order to minimize this angular effect, we have chosen the width $\ell$ of each layer to be of the order of a mean free path: $\ell=0.5$. 

The total width of the system is given by $2X = 4\ell M$, where the integer $M$ (the total length of each material in the system) satisfies
\bala
M = \f{1}{\varepsilon}.
\nala
Vacuum boundary conditions are assigned at $x=\pm X$. 

Cross sections and sources in the void material are 0; the parameters of the solid material are
\bala
\Sigma_{t1} &= 1,\\
Q_1 &= \f{0.2}{M^2},
\nala
with the absorption ratio given by
\bala
1-c = \f{0.1}{M^2}.
\nala
As $M$ increases, $\varepsilon$ decreases, and the 1-D system approaches the diffusive limit described in the asymptotic analysis.

To generate benchmark results for comparison, we used the same procedure presented in \cite{vas16}.
In this procedure, we obtain a physical realization of the system by choosing a continuous segment of two full layers (one of each material) and randomly placing the coordinate $x = 0$ in this segment.
Given this fixed realization of the system, the cross sections and source are now deterministic functions of space.

We solve the transport equation numerically for this realization using (i) the standard discrete ordinate method with a 16-point Gauss-Legendre quadrature set (S$_{16}$); and (ii) diamond differencing for the spatial discretization.
This procedure is repeated for different realizations of the random system.
Finally, we calculate the ensemble-averaged scalar flux by averaging the resulting scalar fluxes over all physical realizations.
\begin{figure}[ht] 
  \centering
  \includegraphics[scale=1.0]{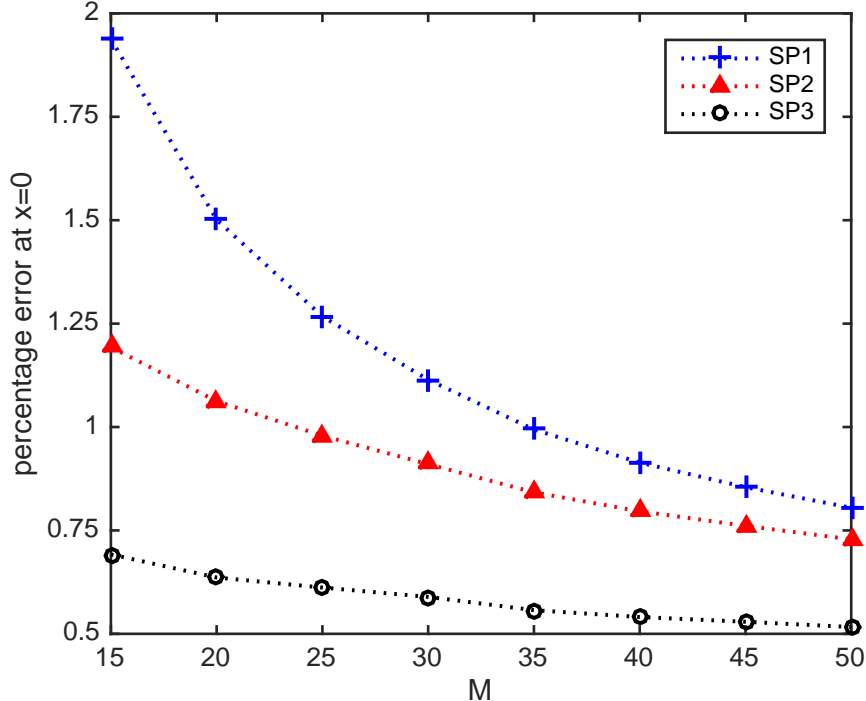}
  \caption{Error of the nonclassical SP$_N$ estimates for the scalar flux with respect to the benchmark solution at $x=0$.}
  \label{fig}
\end{figure}

The expectation is that the estimates obtained with the nonclassical SP$_N$ theory as given by \cref{29} to (\ref{32}) will increase in accuracy as $M$ increases and $\ep\rightarrow 0$.
\Cref{fig} shows the percent relative error of the nonclassical SP$_N$ estimates with respect to the benchmark results, calculated at the center of the system ($x=0$) for different values of $M$. 

As anticipated, the numerical results confirm the asymptotic analysis: (i) the accuracy of the SP$_N$ equations increases as $N$ increases; and (ii) for each choice of $N$, the error decreases as $M$ increases and the system approaches the diffusive limit.

\section{Conclusions}
In this paper we have derived a set of diffusion approximations to the nonclassical transport equation with isotropic scattering using a high-order asymptotic expansion.
These approximations reduce to the simplified P$_N$ equations under the assumption of classical transport, and for that reason are labeled \textit{nonclassical} SP$_N$ equations.
Explicit equations are given for nonclassical SP$_1$ (diffusion), SP$_2$, and SP$_3$; higher-order equations can be derived by continuing to follow the same procedure.
The caveat of this analysis is that the first $2M$ raw moments of the free-path distribution $p(s)$ are required to be finite in order to obtain the nonclassical SP$_N$ equations for $N=M$.

Although the analysis does not yield boundary conditions, we show that the nonclassical SP$_N$ equations can be manipulated into a classical form with modified parameters.
This allows us to generate numerical results using Marshak vacuum boundary conditions.
More importantly, by using this approach one can implement the nonclassical SP$_N$ equations in existing SP$_N$ codes.

Numerical results for a 1-D random periodic system are presented, validating the theoretical predictions.
This result paves the road to a more complete understanding of the diffusive behavior of the nonclassical transport theory.
Future work includes (i) performing numerical calculations in nonclassical multi-dimensional systems; (ii) extending the asymptotic analysis to include angular-dependent free-path distributions; and (iii) extending the analysis to include anisotropic scattering.

\section{Acknowledgments}
This paper was prepared by R. Vasques and R.~N. Slaybaugh under award number NRC-HQ-84-14-G-0052 from the Nuclear Regulatory Commission.
The statements, findings, conclusions, and recommendations are those of the authors and do not necessarily reflect the view of the U.S. Nuclear Regulatory Commission.

\bibliographystyle{ans}
\bibliography{bibliography}
\end{document}